# La antimateria

## Beatriz Gato Rivera

La antimateria es uno de los aspectos más fascinantes de la Física de Partículas y también uno de los más desconocidos, a pesar de vivir rodeados por ella y por las radiaciones resultantes de su aniquilación contra la materia.

La antimateria, el reverso de la materia, es uno de los aspectos más fascinantes a la vez que más desconocidos de la Física de Partículas. Curiosamente, un objeto de antimateria sería indistinguible, a juzgar por su aspecto, de uno de materia; de hecho, si existieran estrellas de antimateria brillarían de manera idéntica que sus homólogas de materia, emitiendo la misma luz. Sin saberlo, vivimos rodeados por antimateria y por las radiaciones resultantes de aniquilarse esta contra la materia de su entorno. Por ejemplo, en la superficie terrestre estamos sometidos a una lluvia incesante de partículas, tanto de materia como de antimateria, producidas en las cascadas de rayos cósmicos; y se estima que un 10 % de la luz visible que nos llega del Sol se debe a la aniquilación materia-antimateria que tiene lugar en su interior, pues los hornos nucleares de las estrellas son grandes productores de antielectrones que se aniquilan inmediatamente con los electrones del plasma en el que están inmersos. Asimismo, en nuestra sociedad hacemos amplio uso de la antimateria, especialmente de los antielectrones, tanto en medicina —siendo el ingrediente esencial de los escaneos PET— como en la industria y en la investigación relacionada con la Ciencia y Tecnología de Materiales.

### 1. ¿Qué es la antimateria?

Para comprender lo que es la antimateria hay que adentrarse en el mundo de las partículas elementales; y desde allí puede decirse que la antimateria es el reverso de la materia porque las partículas y sus antipartículas tienen propiedades opuestas, además de algunas propiedades idénticas que no admiten valores opuestos. Así, las partículas y sus antipartículas tienen la misma masa, el mismo espín y la misma vida media, pero valores opuestos de la carga fuerte, la carga débil, la carga eléctrica, el número bariónico, el número leptónico y la helicidad. Todas las partículas tienen antipartículas, aunque en algunos casos coinciden entre sí, como ocurre con el fotón y el bosón de Higgs.

El Modelo Estándar que describe las partículas elementales está basado en la Teoría Cuántica de Campos, que aúna la mecánica cuántica con la relatividad especial para describir las partículas a grandes velocidades. Según esta teoría, las partículas elementales son solo excitaciones de los campos cuánticos, que llenan todo el espacio, todo el Universo; y cada especie de partícula tiene su propio campo, con la salvedad de que una partícula y su antipartícula comparten el mismo. No es de extrañar, por tanto, que a menudo se produzcan pares partícula-antipartícula al excitar los campos cuánticos con la energía de las colisiones, ya sea en los aceleradores de partículas o en la atmósfera por la acción de los rayos cósmicos.

La primera antipartícula encontrada, el antielectrón, fue denominada *positrón* por su descubridor oficial, Carl Anderson, un joven investigador en Caltech, California, que investigaba las trazas que dejaban las partículas de la radiación cósmica en la pequeña cámara de niebla de su laboratorio. El hallazgo se publicó en 1932 y, curiosamente, otros científicos se percataron de que también habían observado esas curiosas trazas, similares a las dejadas por los electrones, pero girando al revés, denotando carga eléctrica positiva. Se da la circunstancia, además, de que la existencia del antielectrón fue propuesta solo dos años antes por el físico teórico Paul Dirac, para dar cuenta de las soluciones de energía negativa de su famosa ecuación, que describe los electrones.

En cuanto a los átomos de antimateria, no se sabe si existen en la naturaleza en algún lugar del Universo; los únicos que hemos conocido fueron creados en laboratorios, entre 1995 y 2018, sobre todo en la Factoría de Antimateria del CERN, donde volverán a producirse a partir de 2021.

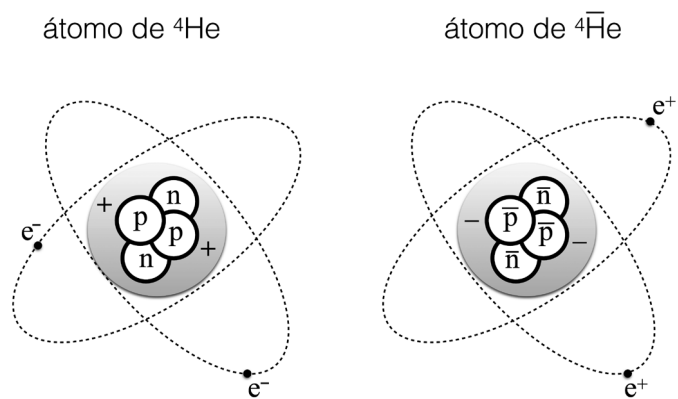

**Fig. 1.** Esquema de un átomo de Helio-4 y un átomo de Antihelio-4

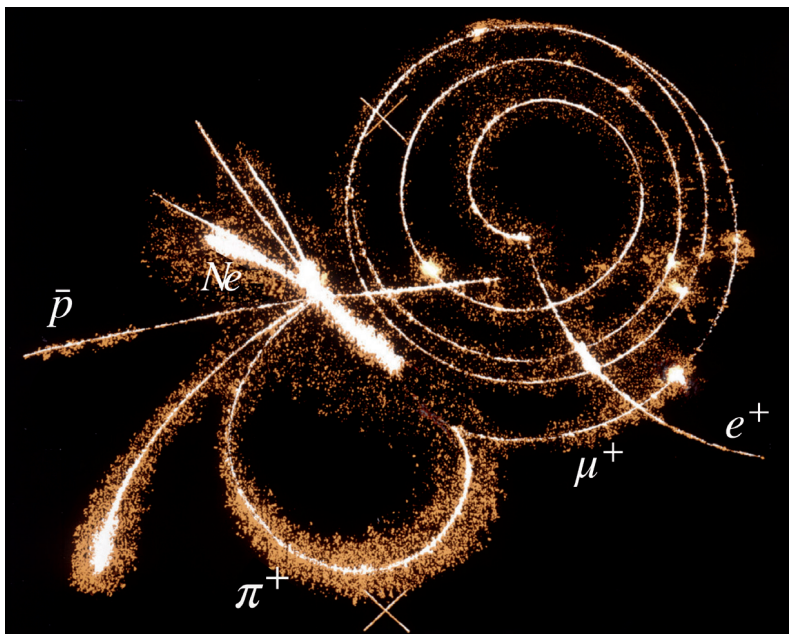

**Fig. 2.** Aniquilación de un antiprotón contra un núcleo de un átomo de neón. Experimento PS-179 en la máquina LEAR del CERN (1984). (Imagen cortesía del CERN).

**Fig. 3.** Esquema de una cascada de rayos cósmicos.

### 2. La aniquilación materia-antimateria

Un rasgo muy distintivo de la antimateria es que al entrar en contacto con la materia ambas se aniquilan entre sí. Si el encuentro entre una partícula y su antipartícula tiene lugar a pequeñas velocidades; es decir, a bajas energías, el resultado de la aniquilación consiste solo en fotones; pero si la colisión ocurre con energías suficientemente altas también se pueden crear otras partículas y antipartículas, además de fotones. Este fenómeno sucede de manera habitual en las colisiones entre partículas, y se debe a la fórmula de Einstein de la conversión entre masa y energía, $E = mc^2$, por la cual parte de la energía de la colisión puede invertirse en crear partículas y antipartículas con masas a su alcance, excitando sus campos cuánticos. Como los fotones no tienen masa, no se requiere un umbral de energía para que estos se produzcan.

También hay que señalar que la aniquilación materia-antimateria es el proceso más energético que existe. Por ejemplo, un gramo de antimateria lanzado sobre nuestro planeta produciría una deflagración equivalente a casi tres veces la bomba atómica que asoló Hiroshima en 1945, suponiendo que esta tuviera 15 kilotones de TNT (se barajan entre 12 y 18). A pesar de ello, no es posible utilizar la aniquilación materia-antimateria para cubrir las necesidades energéticas de nuestra sociedad, como hacemos con la energía nuclear; de hecho, estamos muy lejos de poder conseguirlo. La razón fundamental es que con la tecnología actual solo es posible producir cantidades irrisorias de antimateria y además a costa de un gigantesco esfuerzo, tanto tecnológico como energético y financiero. Por ejemplo, la mayor factoría de antiprotones que ha existido, encargada de suministrarlos al acelerador Tevatrón, del Fermi National Laboratory (Fermilab, EE. UU.) para hacerlos colisionar contra protones, logró producir unos $10^{15}$ antiprotones por año, apenas suficientes para llevar a ebullición un litro de agua a temperatura ambiente al aniquilarse con el mismo número de protones (se necesitarían $1,4 \times 10^{15}$ aniquilaciones $p\bar{p}$, para elevar la temperatura de un litro de agua desde 0 °C hasta 100 °C). Y por si esto fuera poco, teniendo en cuenta que un gramo de antiprotones equivale a $6,02 \times 10^{23}$ de los mismos, serían necesarios 602 millones de años para producir un solo gramo usando la tecnología de la factoría de Fermilab. En cuanto al CERN, usando su tecnología actual se necesitarían 60.200 millones de años para producir un gramo de antiprotones, más de cuatro veces los 13.800 millones de años que lleva existiendo el Universo.

### 3. Fuentes de antimateria

Es crucial diferenciar entre la antimateria primordial y la antimateria secundaria. La primera se creó al comienzo del Universo, en los primeros instantes después del Big Bang, hace unos 13.800 millones de años, y bien pudiera haber desaparecido en su práctica totalidad. La antimateria secundaria, por el contrario, se está creando continuamente en nuestro entorno, ya sea por colisiones entre partículas —como sucede en las cascadas de rayos cósmicos— o en las reacciones nucleares dentro de las estrellas y otros procesos astrofísicos, o bien en los núcleos de ciertas sustancias radiactivas, como el isótopo de potasio $^{40}$K, que abunda en los plátanos, en las paredes y en nuestros propios huesos. De hecho, este isótopo emite el 89,28 % de las veces un electrón más un antineutrino, es decir, radiación $\beta$; el 10,72 % de las veces emite radiación gamma más un neutrino; y el 0,001 % de las veces emite un positrón más un neutrino, la llamada radiación $\beta^+$. Los positrones emitidos por los huesos no llegan a salir del cuerpo, pues se aniquilan con los electrones a su paso y su recorrido es de apenas un milímetro. En el caso de los plátanos, como también tienen $^{40}K$ en la piel, escapan al exterior unos 15 positrones cada 24 horas. Así que los plátanos emiten antimateria, y lo mismo puede decirse de la cerámica, las paredes y similares, ya que la cal y el cemento también contienen $^{40}K$. Esto nos crea la duda de si el primer positrón detectado por Carl Anderson provenía de una cascada de rayos cósmicos, producido en la desintegración de un antimuón que atravesó el techo de su laboratorio, o provenía de las paredes de este, o de un plátano que se había traído para almorzar.

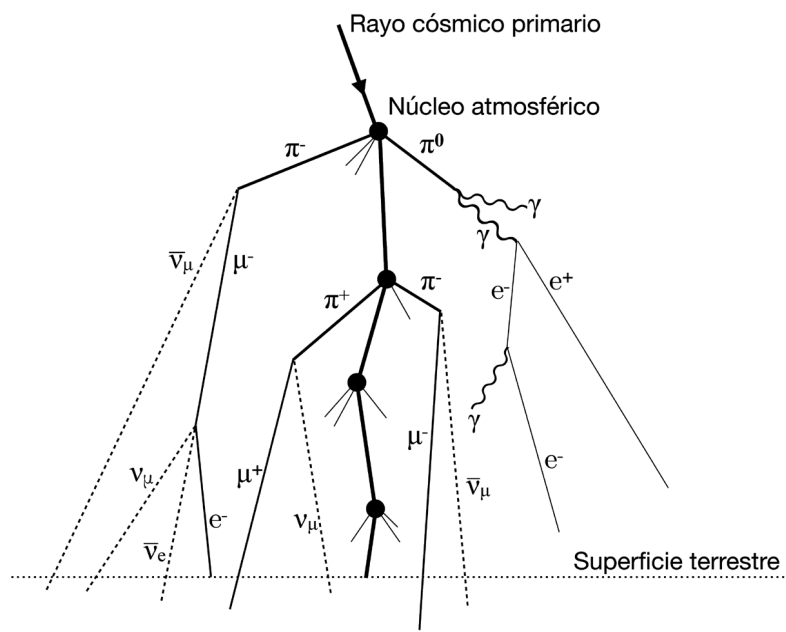

Las fuentes principales de antimateria a nuestro alcance son los aceleradores de partículas y nuestra propia atmósfera, donde chocan los rayos cósmicos con los átomos allí presentes produciendo grandes cascadas de radiación secundaria, tanto mayores cuanto más energéticos sean los rayos primarios. Sorprendentemente, los rayos cósmicos pueden alcanzar energías varios órdenes de magnitud superiores a las más altas que se alcanzan en los aceleradores; y como decíamos, en ambos casos parte de la energía que se produce en las colisiones se invierte en crear otras partículas, tanto de materia como de antimateria (mitad y mitad, aproximadamente).

Las cascadas de rayos cósmicos pueden estar formadas por miles de millones de partículas, muchas de las cuales alcanzan la superficie terrestre. Por ejemplo, a nivel del mar se estima que cada minuto llega un muón (o antimuón) por cm$^2$, y también llegan, entre 10 y 100 veces más, tanto electrones como positrones, además de miríadas de neutrinos, antineutrinos y fotones. Los muones y antimuones más energéticos incluso penetran en nuestras casas y pueden atravesar kilómetros de tierra antes de desintegrarse, mientras que los neutrinos y antineutrinos atraviesan el planeta entero sin apenas inmutarse. Por otro lado, la radiación secundaria también puede salir de la atmósfera hacia afuera y, si se trata de partículas con carga eléctrica, estas pueden quedarse atrapadas en el campo magnético terrestre formando parte de los cinturones de radiación de Van Allen. A este respecto, hay que destacar que en 2011 el detector PAMELA, a bordo del satélite ruso Resurs-DK1, encontró un subcinturón de antiprotones en el cinturón interior, producidos al colisionar rayos cósmicos muy energéticos con moléculas de las altas capas de la atmósfera. El detector constató, además, que su flujo excede en un factor 1.000 al de los antiprotones que viajan libremente por el espacio, por lo que este subcinturón de Van Allen constituye la fuente más abundante de antiprotones cerca de la Tierra.

Además de las reacciones nucleares en las estrellas, existen otros muchos procesos astrofísicos que también producen antimateria, e incluso la liberan al espacio exterior, por lo que una pequeña fracción de los rayos cósmicos primarios, menos del 1 %, consiste en antiprotones y positrones (el 98 % son protones y partículas alfa, y el 2 % restante consiste principalmente en núcleos atómicos). A este respecto, desde 2006 se está intentando capturar en el espacio exterior núcleos atómicos de antimateria mayores que un antiprotón, como las partículas antialfa, consistentes en dos antiprotones y dos antineutrones. Primero se lanzó el detector PAMELA, cuya misión finalizó en 2016, y desde 2011 se encuentra operando el detector AMS-02, instalado en la Estación Espacial Internacional. Pero de momento no se ha tenido éxito con esa búsqueda.

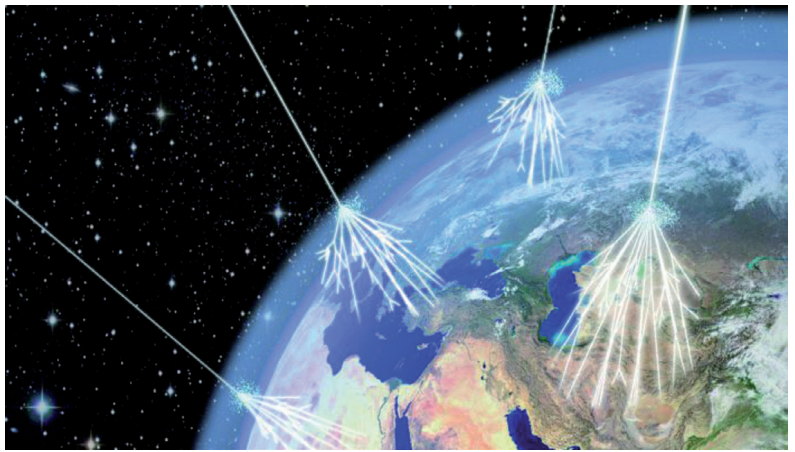

### 4. Experimentos con antiátomos

En 1995 un equipo de científicos del CERN logró producir los primeros antiátomos, un total de nueve átomos de antihidrógeno $\bar{H}$, constituidos por un solo antiprotón en el núcleo y un positrón en la corteza. Dos años después, Fermilab anunciaba que había logrado producir otros 100 antiátomos, pero en ambos casos estos se movían muy rápido y se aniquilaban con la materia circundante antes de poderlos analizar. Esta situación cambió en el año 2000, al entrar en funcionamiento el desacelerador de antiprotones AD (Antiproton Decele-rator) del CERN, que permite producir átomos de $\bar{H}$ más lentos. Lo instalaron en una gran sala junto con varios experimentos, a los que suministraba antiprotones para crear átomos de $\bar{H}$ y analizarlos. Estos experimentos, con los nombres: ATRAP, ALPHA, ASACUSA y AEGIS, se interrumpieron en septiembre de 2018, y continuarán en 2021 cuando todos los aceleradores del CERN reanuden su funcionamiento después de la parada técnica actual, mientras se procede a las mejoras del gran acelerador LHC para su tercera y última fase. Entonces también contarán con el nuevo experimento GBAR y el nuevo desacelerador ELENA, que se acoplará al desacelerador AD y permitirá reducir la velocidad de los antiprotones 50 veces más. Para ser precisos, ELENA será capaz de frenar los antiprotones provenientes del desacelerador AD con una energía de 5,3 MeV, hasta una energía de 0,1 MeV.

**Fig. 4.** Recreación artística dando una perspectiva de la llegada de rayos cósmicos a la Tierra y la formación de cascadas de radiación secundaria. (Crédito: Asimmetrie/INFN).

**Fig. 5.** Vista aérea del complejo de aceleradores del Fermi National Accelerator Laboratory (Fermilab), a 50 km de Chicago. El anillo en segundo plano es el Tevatron, de casi 7 km de longitud, un colisionador protón-antiprotón que fue el acelerador más potente del mundo entre 1985 y 2009, clausurado en 2011. (Fotografía cortesía de Fermilab).

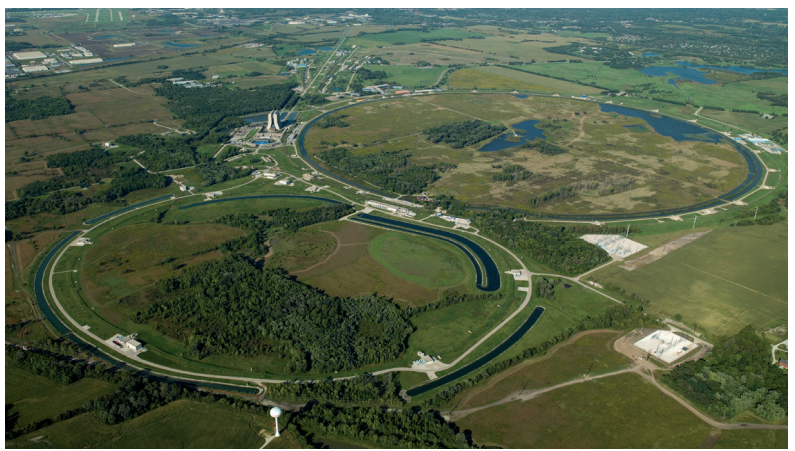

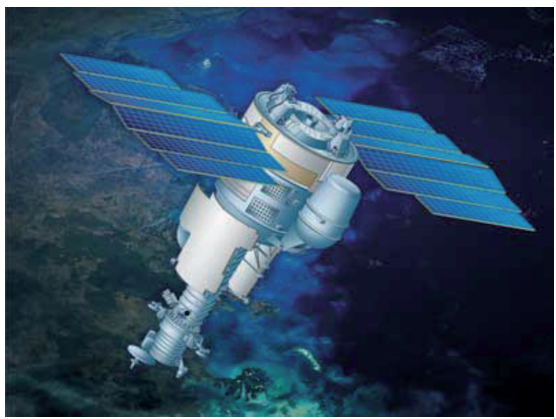

**Fig. 6.** El detector de rayos cósmicos PAMELA a bordo del satélite ruso Resurs-DK1. Entre sus muchos logros, en 2011 descubrió un subcinturón de antiprotones en el cinturón interior de Van Allen. (Imagen cortesía de la colaboración PAMELA).

En los experimentos que se han realizado hasta el presente se han utilizado sobre todo técnicas de espectroscopía, láser o de microondas, para comprobar si los niveles de energía de los átomos de $\bar{H}$ son idénticos a los de los átomos de $H$, a juzgar por los espectros de emisión y absorción de fotones; y hasta la fecha no se ha observado ninguna diferencia entre ellos.

La colaboración ALPHA es la que lleva la delantera en estas investigaciones. En diciembre de 2016 publicó la primera observación de una línea espectral en átomos de antimateria, la correspondiente a la transición 1S-2S en átomos de $\bar{H}$. El método seguido consistió en irradiar los antiátomos confinados en una trampa magnética con dos haces enfrentados de luz láser ultravioleta de longitud de onda λ = 243 nm, la cual es resonante con la transición 1S-2S en átomos de hidrógeno, pues la absorción de dos de estos fotones enfrentados hace saltar al electrón del orbital 1S al orbital 2S. Así, si los átomos de $\bar{H}$ tuvieran exactamente los mismos niveles energéticos que los átomos de $H$, esta técnica haría que gran parte de los positrones que se encontrasen en su estado fundamental en el orbital 1S se excitasen saltando al orbital 2S, como así sucedió. Poco después, en 2018, la colaboración ALPHA publicó la primera observación de la línea espectral Lyman-α, con λ = 121,57 nm, en los átomos de $\bar{H}$, correspondiente a la transición 1S-2P; pero un año antes ya había logrado inducir transiciones atómicas hiperfinas dentro del nivel fundamental 1S exponiendo los átomos de $\bar{H}$ a una radiación de microondas. Estos subniveles se originan por la interacción entre el espín del positrón y el espín del antiprotón en el núcleo atómico. Les siguieron otras transiciones hiperfinas, dentro del nivel 2P, y con los resultados de todas estas observaciones el equipo pudo deducir para los átomos de $\bar{H}$ el homólogo del efecto Lamb, que es la diferencia de energía entre los niveles 2S y 2P con momento angular total $j = 1/2$. Estos niveles están separados por una energía diminuta, de $4,372 \times 10^{-6}$ eV, y el efecto se debe a la interacción de los electrones, o positrones, con las fluctuaciones cuánticas del campo electromagnético en el vacío, tal como predice la electrodinámica cuántica. Este resultado acaba de publicarse en *Nature*, en febrero de 2020.

Hay que destacar también que el experimento ASACUSA crea e investiga helio antiprotónico, en el que un electrón del átomo de helio es sustituido por un antiprotón, produciéndose así las llamadas "atómculas", híbridas compuestas de materia y de antimateria, que se comportan como un átomo y una molécula a la vez.

Sorprendentemente, también se intenta dilucidar si los átomos de $\bar{H}$ caen en el campo gravitatorio terrestre con la misma aceleración que la materia; es decir, si la antimateria participa de la universalidad de la caída libre. No se han obtenido resultados todavía, pero no faltan quienes apuestan por la posibilidad de que esta sienta repulsión gravitatoria hacia la materia, por lo que los átomos de $\bar{H}$ en vez de caer se elevarían en el campo gravitatorio terrestre.

### átomo de helio antiprotónico

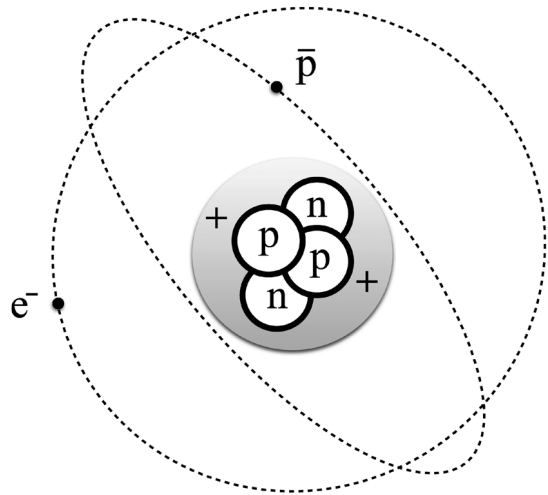

**Fig. 7.** Esquema de una atómcula de helio antiprotónico.

## 5. La antimateria primordial

La antimateria primordial podría haberse aniquilado completamente contra la materia al comienzo del Universo; pero de existir en la actualidad, podría haber dado lugar a estructuras tales como antiestrellas, antiplanetas e incluso pequeñas antigalaxias. Las antiestrellas no se distinguirían de las estrellas ordinarias a juzgar por su aspecto, pues emitirían el mismo espectro electromagnético, pero podríamos diferenciarlas porque las estrellas que consumen hidrógeno, como el Sol, emiten neutrinos, mientras que sus análogas de antimateria, emitirían antineutrinos. Ahora bien, como aún no tenemos instrumentos que nos permitan dilucidar si una estrella emite neutrinos o antineutrinos, a excepción del Sol, habrá que esperar a futuras tecnologías que puedan hacerlo. No obstante, hay varios tipos de observaciones astronómicas que nos pueden dar información sobre si existen estructuras de antimateria. Estas observaciones se encuadran en tres frentes: el fondo cósmico de radiación difusa de rayos gamma, DGRB; la radiación cósmica de fondo de microondas, CMB, que se liberó cuando se formaron los primeros átomos, 380.000 años después del Big Bang; y los rayos cósmicos en el espacio.

Las radiaciones DGRB y CMB podrían revelar fronteras de aniquilación entre dominios de materia y dominios de antimateria, de las que aún no se tiene constancia. Claro está que la imposibilidad de detectar dichas fronteras no significa que se deba descartar tal posibilidad, y las no-observaciones sirven para obtener cotas sobre la abundancia de la antimateria primordial frente a la materia, que debe ser menor que un antiprotón entre decenas de millones de protones. Así que en la actualidad no puede descartarse que existan pequeñas islas de antimateria por el Universo, incluso dentro de nuestra propia galaxia, que escaparían a la detección.

Los rayos cósmicos, por su parte, podrían dar pruebas irrefutables de la existencia de antiestrellas si se descubriesen núcleos de anticarbono, o mayores, que solo podrían haberse fraguado en reacciones nucleares dentro de estas, del mismo modo que los núcleos de carbono solo se pudieron crear a partir de reacciones nucleares en las estrellas ordinarias. La detección de núcleos de antihelio, sin embargo, ni siquiera probaría que existe antimateria primordial, pues estos núcleos podrían haberse formado en procesos astrofísicos conocidos muy energéticos. Al menos esto es lo que piensan muchos expertos, aunque hay cierta discrepancia entre ellos.

## 6. La asimetría materia-antimateria

El problema de la asimetría materia-antimateria en el Universo radica en la suposición de que las partículas y sus antipartículas se crearon en cantidades idénticas en los primeros instantes después del Big Bang (o, si no lo eran, la inflación cosmológica se encargó de hacerlas idénticas). Pues si se crearon en cantidades iguales en un Universo diminuto y extremadamente denso y caliente, deberían haberse aniquilado entre sí en su práctica totalidad. Así que, ¿cómo sobrevivió la materia a esa Gran Aniquilación que terminó prácticamente con la antimateria? En realidad, la Gran Aniquilación casi termina también con la materia pues, un segundo después del Big Bang, por cada protón primordial que sobrevivió miles de millones sucumbieron a la extinción, y ese protón quedó inmerso en un baño de miles de millones de fotones que, 380.000 años después, dieron lugar a la radiación de fondo de microondas CMB que observamos en la actualidad. Pero se piensa que algo tuvo que suceder antes de la Gran Aniquilación para que se generase un ligerísimo excedente de partículas sobre antipartículas, el cual bastó para que el Universo material llegara a existir tal como lo conocemos.

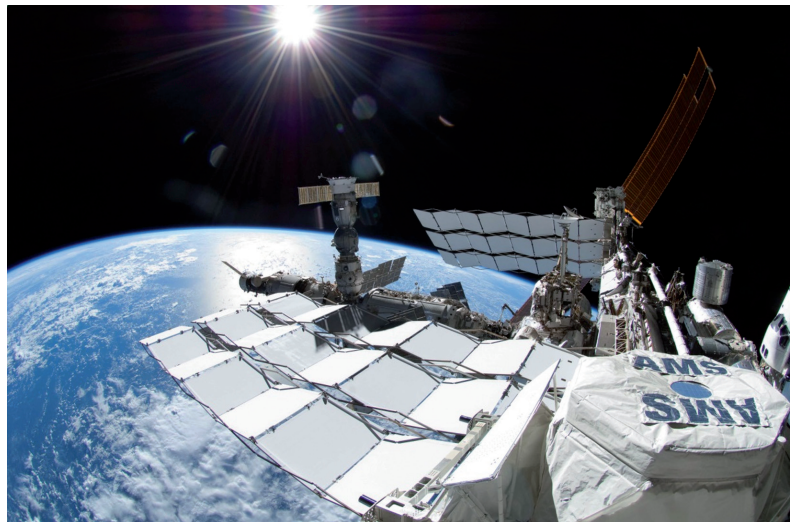

**Fig. 8.** Espectrómetro Magnético Alpha (AMS-02), detector de rayos cósmicos instalado en la Estación Espacial Internacional. (Fotografía cortesía de la NASA).

De hecho, las partículas y sus antipartículas no se comportan siempre de manera idéntica (aparte de que tengan propiedades opuestas). Por ejemplo, en algunos casos se desintegran con diferentes ritmos en canales de desintegración específicos. Esto sucede con el leptón tau, los mesones $K^0$ y todos los mesones de tipo B (los mesones están constituidos por un quark y un antiquark). Este fenómeno se conoce como violación de la simetría CP (carga-paridad), simetría que intercambia partículas y antipartículas entre sí, y cuya violación constituye una de las tres condiciones que Andrei Sájarov propuso para que los procesos entre partículas puedan generar un excedente de materia sobre antimateria. Las otras dos condiciones son la violación del número bariónico B y que los procesos ocurran fuera del equilibrio térmico; es decir, que sean irreversibles. Estas tres condiciones se cumplieron durante la expansión primordial del Universo, pero analizándolas a la luz del Modelo Estándar se concluye que no pudieron producir la cantidad de materia que se observa, sino cantidades varios órdenes de magnitud menores. En consecuencia, la creación de protones, neutrones y electrones primordiales (bariogénesis y leptogénesis), constituye uno de los problemas más acuciantes de la Física de Partículas en la actualidad.

**Fig. 9.** Escaneos PET del cerebro de una persona alcohólica, 10 días (izqda) y 30 días (dcha) después de comenzar la cura de abstinencia. (Crédito: MacGill University, Canadá).

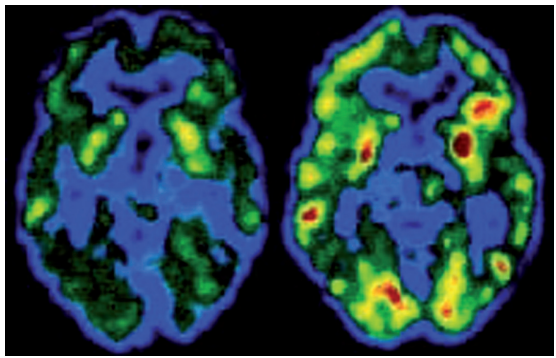

### 7. Aplicaciones médicas y tecnológicas de la antimateria

La aniquilación de los electrones de la materia biológica con positrones provee una de las técnicas de imagen más importantes que se usan en los hospitales. Se trata de la tomografía por emisión de positrones (PET), en la cual se inyectan en el paciente fármacos con isótopos radiactivos de vida muy corta que emiten positrones. Estos se aniquilan con los electrones del tejido circundante emitiendo dos rayos gamma de 511 keV en direcciones opuestas, que son detectados por el escáner PET, dentro del cual se aloja el paciente. De este modo, el aparato va confeccionando una serie de láminas a través del cuerpo que se combinan formando una imagen en 3-D. El escáner PET puede seguir procesos biológicos tales como el metabolismo, la transmisión neuronal y el crecimiento de tumores, por lo que está muy indicado para estudiar el comportamiento de medicamentos en el cuerpo, investigar la actividad del cerebro y obtener imágenes precisas de tumores en 3-D.

En el futuro se usarán también antiprotones, en lugar de protones, en la radioterapia contra tumores, lo que permitirá disminuir sustancialmente el nivel de radiación al paciente debido al mayor poder destructivo de los antiprotones al alcanzar las células tumorales.

En cuanto a las aplicaciones tecnológicas de la antimateria, la radiación con positrones (y en algunos casos antimuones) se usa ampliamente para estudiar materiales con propiedades tecnológicas importantes, como ferroelectricidad, superconductividad y magnetoresistencia, así como materiales semiconductores utilizados en paneles solares y otros dispositivos electrónicos. También se usa para obtener imágenes del interior de algunos materiales con finalidades varias, desde analizar procesos de interés industrial (residuos nucleares, fluir del aceite, etc.), hasta encontrar defectos estructurales en cristales o sondear escalas subnanométricas en biopolímeros.

### Lecturas recomendadas

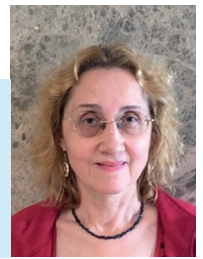

**Beatriz Gato Rivera**
Instituto de Física Fundamental, CSIC